\documentclass[10pt, preprintnumbers]{iopart}

\usepackage{iopams,setstack} 
\usepackage{cite} 
\usepackage{graphicx}
\usepackage{bm}
\usepackage{amssymb} 
\usepackage{hyperref} 
\usepackage{csvsimple-l3, booktabs} 
\usepackage[caption=false]{subfig} 
\usepackage[margin=30mm]{geometry}
\usepackage{xcolor} 
\usepackage[british]{babel}
\hypersetup{colorlinks=true, anchorcolor=black} 

\newcommand{\vect}[1]{\bm{#1}}
\newcommand{\transpose}{\intercal}

\newcommand{\preprint}{
  \setlength{\unitlength}{1mm}{\hbox{\begin{picture}(0,0)
        \put(100,10){\mbox{\footnotesize%
            ADP-24-21/T1260}}\end{picture}}}}

\begin{document}

\title[]{\preprint
Physical interpretation of the $\mathbf{2s}$ excitation of the nucleon}
\author{
Finn~M.~Stokes,
Waseem~Kamleh,
Derek~B.~Leinweber, 
Benjamin~J.~Owen\footnote{Now at the Bureau of Meteorology, Adelaide, SA, Australia.}
}
\address{ARC Special Research Centre for the Subatomic Structure of Matter (CSSM),
         Department of Physics, The University of Adelaide, SA, 5005, Australia.}
\ead{finn.stokes@adelaide.edu.au}
	
\begin{abstract}
Lattice QCD calculations of the $2s$ radial excitation of the nucleon place the state at an energy
of approximately 1.9 GeV, raising the possibility that it is associated with the $N1/2^+(1880)$ and
$N1/2^+(1710)$ resonances through mixing with two-particle meson-baryon states.  The discovery of
the $N1/2^+(1880)$ resonance in pion photoproduction but not in $\pi N$ scattering and the small
width of the $N1/2^+(1710)$ resonance suggest that a state associated with these resonances would
be insensitive to the manner in which pions are permitted to dress it.
To explore this possibility, we examine the spectrum of nucleon radial excitations in both
2+1 flavour QCD and in simulations where the coupling to meson-baryon states is significantly modified
through quenching.  We find the energy of the $2s$ radial excitation to be insensitive to this modification
for quark masses close to the physical point.  This invariance provides further evidence
that the $2s$ radial excitation of the nucleon is associated with the $N1/2^+(1880)$ and
$N1/2^+(1710)$ resonances.
\end{abstract}

\noindent Keywords: lattice QCD, hadron spectroscopy, variational method, radial excitations

	
	
\section{Introduction}
\label{sec:intro}

\subsection{First radial excitation of the nucleon in lattice QCD}

The realisation that the Roper resonance is not the anticipated $2s$ radial excitation of a
constituent quark model was established almost a decade ago \cite{Leinweber:2015kyz} as it became
apparent that the Roper is dynamically generated through rescattering in the $\pi N$, $\pi \Delta$,
and $\sigma N$ channels \cite{Liu:2016uzk}.  This conclusion is driven by lattice QCD calculations
showing the $2s$ radial excitation of the nucleon lies at approximately 1.9 GeV
\cite{Mahbub:2010v1,Mahbub:2010rm,Alexandrou:2014mka,Mahbub:2013ala,%
  Leinweber:2015kyz,Roberts:2013ipa,Roberts:2013oea,Kiratidis:2015vpa,%
  Kiratidis:2016hda,Khan:2020ahz,Liu:2016uzk,Lang:2016hnn,Stokes:2019zdd,Hockley:2023yzn}, far from
the Roper resonance of the nucleon at 1.44 GeV \cite{Roper:1964zza}.

The first hint of a surprise in the $2s$ excitation energy was reported in
Ref.~\cite{Mahbub:2010v1}.
By combining narrow and wide Gaussian-smeared sources with opposite signs, a node in the wave
function of the quark distributions could be created.  The generalised eigenvalue solution
\cite{Michael:1985ne,Luscher:1990ck} determined the superposition of Gaussian-smeared quark sources
and the $2s$ radial excitation of the nucleon was observed at 1.79(8) GeV in a lattice volume of 3
fm on a side \cite{Mahbub:2010rm}. 

These first results were confirmed by the HSC collaboration at heavy quark masses
\cite{Edwards:2011jj} and in an approach using the Athens Model Independent Analysis Scheme
\cite{Alexandrou:2014mka} to extract the spectrum from lattice correlation functions. A
comprehensive analysis of the structure and flow of the states as a function of quark mass followed
in Ref.~\cite{Mahbub:2013ala}.
The wave functions of both the $2s$ and $3s$ excitations were calculated in lattice QCD
\cite{Roberts:2013ipa,Roberts:2013oea} and their profiles were found to compare well with
constituent quark model expectations \cite{Roberts:2013ipa}.

Next generation calculations appeared in 2016 with a high-statistics calculation coming from the
CSSM \cite{Liu:2016uzk} placing the $2s$ state at 1.90(6) GeV on the 3 fm lattice.  More exotic
five-quark descriptions for the Roper resonance were pursued
\cite{Kiratidis:2015vpa,Kiratidis:2016hda}.  However, no new low-lying states have been observed.
Similarly, the consideration of hybrid baryons \cite{Khan:2020ahz} has not revealed any new
low-lying states in the Roper region.

This situation contrasts the case of the $\Lambda(1405)$ channel.  There, three-quark operators
are able to excite a finite-volume state in the resonance region of $\sim 1.4$ GeV.  Lattice QCD
calculations of the electromagnetic form factors of this state show a localised state similar in
size to the ground-state $\Lambda$ baryon \cite{Menadue:2013xqa}.  However, it is the quark mass
dependence of the magnetic form factor that signals the presence of a molecular meson-baryon bound
state in the finite-volume of the lattice \cite{Hall:2014uca,Hall:2016kou}.

%
%

\subsection{Other evidence supporting the \texorpdfstring{$2s$}{2s} excitation at \texorpdfstring{$\sim 2$}{\texttildelow2} GeV}

The very first insights into the role of two-particle scattering states in the Roper resonance
region appeared from Lang {\it et al.} \cite{Lang:2016hnn}.  Scattering states very near the
non-interacting two particle energies were observed.  While not manifesting any obvious resonance
signature, an HEFT analysis bringing experimental phase shifts and inelasticities to the finite
volume of the lattice \cite{Wu:2017qve} illustrates how Lang {\it et al.}'s results are in accord
with the scattering phase-shifts and inelasticities in the Roper resonance region.  The most recent
calculations of the nucleon spectrum are focused on precision calculations of the very lowest-lying
scattering states \cite{Bulava:2023uma,Morningstar:2021ewk,Bulava:2022vpq}.

In addition, the three-quark single particle state was observed in Lang {\it et al.}'s analysis.
It's position persists at the order of 2 GeV, even when lower-lying scattering states are included
in the correlation matrix, thus confirming earlier observations
\cite{Mahbub:2010rm,Mahbub:2013ala,Alexandrou:2014mka,%
  Leinweber:2015kyz,Roberts:2013ipa,Roberts:2013oea,Kiratidis:2015vpa,Kiratidis:2016hda},

The role of chiral symmetry in lattice fermion actions has also been explored quantitatively in
Ref.~\cite{Virgili:2019shg} to see if the explicit breaking of chiral symmetry in Wilson-clover
fermion actions is responsible for the large $2s$ excitation energy.  A direct analysis of lattice
correlation functions from Wilson-clover and overlap fermion actions -- the latter providing a
lattice implementation of chiral symmetry -- reveals no differences in the spectrum.

Since then, parity-expanded variational analysis (PEVA) techniques \cite{Stokes:2013fgw} have been
developed to explore the electromagnetic form factors of both the even and odd-parity excitations
of the nucleon \cite{Stokes:2019zdd} and their electromagnetic transitions to the ground state
\cite{Stokes:2019yiz}.  With regard to the $2s$ excitation, the charge radius of the excited proton
is larger than the ground state, in accord with expectations.  Moreover, the magnetic moments of
the excited $2s$ proton and neutron calculated in lattice QCD agree with the ground-state magnetic
moments, again consistent with the expectations of a $2s$ excitation.

In summary, the $2s$ radial excitation of the nucleon observed in lattice QCD is firmly entrenched
at approximately 2 GeV. The most precise determination \cite{Liu:2016uzk} places it at 1.90(6) GeV
in a 3 fm lattice volume. The single-particle basis state contribution in the Roper resonance
energy regime is a few percent at best \cite{Wu:2017qve}.  In the literature, the $2s$ quark-model
excitation is often referred to as the Roper state.  We now know this is not the case. But then,
what state(s) in the nucleon spectrum are associated with the $2s$ single-particle excitation of
the nucleon?

\subsection{\texorpdfstring{$N1/2^+(1710)$}{N1/2+(1710)} and \texorpdfstring{$N1/2^+(1880)$}{N1/2+(1880)} resonances}

With the $2s$ radial excitation at $\sim 2$ GeV near the physical point,
we turn our attention to nucleon resonances in this energy regime.  As the single-particle quark
model state mixes with nearby two-particle states to form resonances, we anticipate the $2s$ radial
excitation is largely associated with the $N1/2^+(1880)$ resonance observed in meson
photoproduction \cite{Anisovich:2017bsk} and to some extent the $N1/2^+(1710)$ as it is only 170
MeV away.  
%

The $N1/2^+(1710)$ resonance has a small width of 80 to 200 MeV \cite{ParticleDataGroup:2024cfk}
relative to the Roper resonance for example at 250 to 450 MeV.  Moreover, the $\pi N$ partial width
of the $N1/2^+(1710)$ is in the range of 5 to 20\%.  Within the finite volume of the lattice, these
resonance properties suggest that the $2s$ excitation may have relatively weak coupling to the
long-range $\pi N$ components.
The $N1/2^+(1880)$ resonance is a curious state that was not seen in $\pi N$ scattering
experiments.  Rather it was discovered through $\gamma p \to K^+ \Lambda$ photoproduction
experiments \cite{Anisovich:2017bsk}.  While its width is estimated at 200 to 400 MeV, the $\pi N$
partial width may be as small as 3\% \cite{ParticleDataGroup:2024cfk}, in accord with its lack of
signature in $\pi N$ scattering. Again, these resonance properties suggest that the $2s$ excitation
observed on the lattice may have relatively weak coupling to the long-range $\pi N$ components.

This admits the possibility of an insensitivity of the $2s$ radial excitation energy to changes to
the manner in which meson-baryon states dress the state observed on the lattice. One could probe
this possibility by altering these dressings and measuring the impact of these alterations on the
observed spectrum.  

In short, if the finite-volume $2s$ state observed in lattice QCD is associated with the
$N1/2^+(1710)$ and $N1/2^+(1880)$ resonances, there is good reason to anticipate that the mass of
the state is insensitive to modifications of the meson cloud dressing.  This possibility is
explored herein.

%
%

\subsection{Modifying the meson-baryon couplings}

Our approach is to significantly modify the strength of meson-baryon two-particle couplings to the
eigenstates of the finite-volume nucleon spectrum and search for the existence of states
insensitive to this variation.  We implement this by removing sea-quark-loops from the gauge field
generation, eliminating a class of quark-flow diagrams and thus suppressing the meson-baryon
couplings.  

The phenomenology of this quenching is well understood through the formalism of (partially)
quenched chiral perturbation theory for baryons \cite{Labrenz:1996jy,Hall:2015cua}.  A
comprehensive analysis of baryon couplings and their modifications under quenching is provided in
Ref.~\cite{Leinweber:2002qb}. The changes are complicated, with some meson dressings changing in
sign and most dressings becoming significantly suppressed. 

For example, consider the pion dressings of the proton. Table 1 of Ref.~\cite{Leinweber:2022guz}
provides the contributions of connected and disconnected quark flows. The disconnected contributions
disappear upon quenching the theory, but the connected contributions remain. The sum of all of these
contributions is 5.38, whereas the connected contributions alone sum to 1.50\footnote{In a scheme where
the physical $n\pi^+$ dressing of the proton is equal to $2{(D+F)}^2=2g_A^2$ with $g_A=1.27$}.
Consequentially, quenching reduces the coupling by a factor of 3.6.

It is well known that ground state hadrons such as the nucleon are approximated well
by quenched QCD \cite{CP-PACS:1999ity,CP-PACS:2002unz}, with a precision of $\sim 1$ to $3$\%
\cite{Fodor:2012gf}.  The success for the lowest-lying hadrons is associated with the dominance of
a single-particle basis state in their structure.
This single-particle dominance is evident in Hamiltonian Effective Field Theory (HEFT).
As illustrated in Fig.~10a of Ref.~\cite{Liu:2016uzk}, HEFT describes
the ground-state nucleon as 78\% single-particle basis state . 

The $2s$ excitation displays a somewhat similar superposition of basis states in HEFT. 
For example, Fig.~5 of Ref.~\cite{Wu:2017qve} illustrates that the single-particle basis state
component of the $2s$ excitation may approach 70\%, with 25\% in the $\pi \Delta$ channel, and only
a few percent in the long-range $\pi N$ and $\sigma N$ channels. 

With 70\% of the state composition insensitive to the meson-baryon basis-state components, one can see
how quenching the theory would generate relatively small changes, particularly if the lattice scale
determination is designed to preserve the physics of full QCD at a physically relevant scale.

To this end, we use the Sommer scale \cite{Sommer:1993ce} which sets the lattice spacing by
preserving the force of the static quark potential at a scale relative to ground-state hadronic
physics.  In this way, the physics at that scale is preserved.  However, changes in the pion-nucleon
couplings to states, for example, mean that the long-distance part of the state is different.

\section{Correlation Matrix Techniques}
\label{CorrelationMatrixTechniques}

To isolate energy eigenstates on the lattice, we use the variational
method~\cite{Michael:1985ne,Luscher:1990ck}, briefly reviewed in the following.
To access $N$ states of the spectrum, one requires a minimum of $N$ interpolators coupling well to
the states of interest. The parity-projected two-point correlation matrix for $\vec{p} =0$ is
introduced as
\begin{eqnarray}
G_{ij}^{\pm}(t) &=& \sum_{\vec x}\, {\rm Tr}_{\rm sp}\, \{
\Gamma_{\pm}\, \langle\, \Omega\, \vert\, \chi_{i}(x)\,
\bar\chi_{j}(0)\, \vert\, \Omega\, \rangle\}, 
\label{eq:CM} \\
          &\approx& \sum_{\alpha = 0}^{N-1}\, \lambda_{i}^{\alpha}\,
\bar\lambda_{j}^{\alpha}\, e^{-m_{\alpha}t},
\end{eqnarray}
where $\chi_{i}$ and $\bar\chi_{j}$ are proton interpolating fields at the sink and source
respectively, $\vert\, \Omega\, \rangle$ denotes the non-trivial ground-state fields,
$\Gamma_{\pm}=(\gamma_{0}\pm 1)/2$ projects the parity of the eigenstates, $\lambda_{i}^{\alpha}$
and $\bar\lambda_{j}^{\alpha}$ are the couplings of the interpolators to the state with mass
$m_\alpha$, and $\alpha$ enumerates the energy eigenstates with the ground state corresponding to
$\alpha = 0$. We use the Pauli representation of the $\gamma$ matrices such that they are
Hermitian.  Dirac indices are implicit.

Using an average of $\{U\} + \{U^*\}$ configurations, our construction of $G_{ij}^{\pm}(t)$ is
symmetric and real.  We enforce this symmetry by working with an improved unbiased symmetric
construction $(G_{ij} + G_{ji})/2$.  To ensure that the matrix elements are $\sim{\cal{O}}(1)$,
each element of $G_{ij}(t)$ is normalised by the diagonal elements of $G_{ij}$ as ${G}_{ij}(t) / (
\sqrt{{G}_{ii}(t_N)}\, \sqrt{{G}_{jj}(t_N)} )$ (no sum on $i$ or $j$) with normalisation time $t_N$
at one time slice after the source.

An operator creating state $\alpha$ can be constructed as
$\bar{\phi}^{\alpha}=\sum_{j}{\bar\chi}_{j}\, u_{j}^{\alpha}$, with linear coefficients $u_j$ for
each state $\alpha$.  As the time dependence of the two-point function is then governed by $\exp(
-m_\alpha\, t)$ a recurrence relation can be used to solve for $u_{j}^{\alpha}$
\begin{equation}
G_{ij}(t_{0}+\triangle t)\, u_{j}^{\alpha} = e^{-m_{\alpha}\triangle
  t}\, G_{ij}(t_{0})\, u_{j}^{\alpha}  \, .
 \label{eq:recurrence_relation}
\end{equation}  
Multiplying from the left by $G^{-1}(t_0)$ provides the right
eigenvector equation for $u_{j}^{\alpha}$
\begin{equation}
[(G(t_{0}))^{-1}\, G(t_{0}+\triangle t)]_{ij}\, u^{\alpha}_{j} =
c^{\alpha}\, u^{\alpha}_{i} \, , 
\label{eq:right_evalue_eq}
\end{equation} 
with $c^{\alpha}=e^{-m_{\alpha}\triangle t}$.  Similarly, an operator
annihilating state $\alpha$ can be defined as $\phi^\alpha =\sum_j
\chi_j\, v_j^\alpha$, where $v_j^\alpha$ is given by the left
eigenvalue equation
\begin{equation}
v^{\alpha}_{i}\, [G(t_{0}+\triangle t)\, (G(t_{0}))^{-1}]_{ij} =
c^{\alpha}v^{\alpha}_{j} \, .
\label{eq:left_evalue_eq}
\end{equation} 
The eigenvectors for state $\alpha$, $u_{j}^{\alpha}$ and $v_{i}^{\alpha}$, are then combined to
create the eigenstate projected correlation function
\begin{equation}
G^{\alpha}_{\pm}(t) \equiv v_{i}^{\alpha}\, G^{\pm}_{ij}(t)\, u_{j}^{\alpha} ,
\label{projected_cf_final}
\end{equation}
with parity $\pm\,$.  We note that with our symmetric construction for 
$G_{ij}^{\pm}(t)$, the left and right eigenvectors are equal.



In exciting the nucleon, we consider the standard local interpolating fields describing the quantum
numbers of the proton
\begin{eqnarray}
  \chi_1(x) &=& \epsilon^{abc} \left ( u^{a\transpose}(x)\, C \gamma_5\, d^b(x) \right )\, u^{c}(x) \,\mbox{, and } \\
  \chi_2(x) &=& \epsilon^{abc} \left ( u^{a\transpose}(x)\, C \,         d^b(x) \right )\, \gamma_5\, u^{c}(x)\, .
\end{eqnarray}
Here $u(x)$ and $d(x)$ represent the Dirac spinor for a single $u$ or $d$ quark at spacetime
position $x$ carrying a colour index $a,\, b,\, c$. The matrix $C$ is the charge conjugation
matrix and $\epsilon$ is the Levi-Cevita antisymmetric tensor.  Gauge-invariant Gaussian
smearing~\cite{Gusken:1989qx} is used to enlarge the basis of operators. Four different smearing
levels are used at the fermion source and sink for each of the two nucleon interpolators, providing
an $8 \times 8$ basis.

The effective mass function is defined as
\begin{equation}
M_{\rm eff}(t) = \frac{1}{n} \log\left ( \frac{G(t)}{G(t+n)} \right )
\, .
\end{equation}
We consider $n=2$ as it provides a more accurate estimation of the local slope of the projected
correlator, while maintaining a good sample of points to identify plateau behaviour.
A second-order single-elimination jackknife analysis \cite{Efron:1979} is used to calculate
uncertainties with the ${\chi^{2}}/{\rm{dof}}$ obtained via the full covariance matrix analysis.

Its well known that scattering states can contaminate the correlation functions of local
single-particle operators. The standard nucleon interpolators used here favour localised states and
tend to miss the non-local scattering states
\cite{Mahbub:2010rm,Mahbub:2012ri,Mahbub:2013ala,Mahbub:2013bba,Kiratidis:2015vpa,Kiratidis:2016hda},
making the lattice spectrum incomplete.  In this case there may be concern that the lattice energy
levels extracted might be systematically contaminated from the missed levels.

The CSSM collaboration has explored this in detail
\cite{Mahbub:2013bba,Kiratidis:2015vpa,Kiratidis:2016hda} and has developed methods that ensure
these systematic errors are suppressed through Euclidean time evolution.  The use of a
single-state Ansatz and a full covariance-matrix calculation of the $\chi^2/{\rm dof}$ with a
conservative cutoff of $\chi^2/{\rm dof} < 1.2$ ensures that any remaining contamination is contained
within the error bars reported \cite{Kiratidis:2016hda}.  


\section{Simulation Details} 
\label{sec:SimDet}

\begin{table}[tb]
\caption{ The relationship between hopping parameter values established in the PACS-CS
  \cite{PACS-CS:2008bkb} dynamical fermion ensembles, $\kappa_D$, and determined in quenched QCD,
  $\kappa_Q$, by matching the pion mass. The quenched ensembles are simulated with the Iwasaki
  action at $\beta = 2.58$ to provide a lattice spacing of $a = 0.100$ fm. The PACS-CS ensembles are
  labelled by the pion mass in the PACS-CS scheme. Lattice spacings for the dynamical ensembles,
  $a_D$, are provided in the Sommer scheme \cite{Sommer:1993ce} with $r_0 = 0.4921(64)(+74)(-2)$ fm.
  The dynamical fermion ensembles contain $n^{\rm cfg}_D$ gauge field configurations with $n^{\rm src}_D$
  sources computed on each configuration, and the quenched ensembles similarly for $n^{\rm cfg}_Q$ and
  $n^{\rm src}_Q$. }
\label{tab:kappas}
\begin{indented}
\item[]\begin{tabular}{@{}cccccccc}
\br 
  $m_\pi/$MeV &$a_D$/fm &$\kappa_D$ &$n^{\rm cfg}_D$ &$n^{\rm src}_D$ &$\kappa_Q$ &$n^{\rm cfg}_Q$ &$n^{\rm src}_Q$ \\
\mr
  701 &0.102 &0.13700 &399 &1 &0.12409 &350 &1 \\
  570 &0.101 &0.13727 &397 &1 &0.12453 &350 &1 \\
  411 &0.096 &0.13754 &449 &2 &0.12495 &350 &1 \\
  296 &0.095 &0.13770 &400 &4 &0.12522 &350 &1 \\
  156 &0.093 &0.13781 &197 &8 &0.12539 &350 &1 \\
%
\br
\end{tabular}
\end{indented}
\end{table}

Our full QCD calculations are performed on the PACS-CS Collaboration's gauge-field ensembles
\cite{PACS-CS:2008bkb} made available via the ILDG \cite{Beckett:2009cb}. These configurations make
use of the Iwasaki gauge action \cite{Iwasaki:1985we} and an $\mathcal{O}(a)$-improved Wilson quark
action with Clover coefficient $C_\mathrm{SW}=1.715$.  The masses of the up and down quarks are
taken to be degenerate.  The simulations use a fixed value of $\beta=1.90$ on a $32^3\times 64$
lattice.
	
Our aim is to replicate the physical aspects of the PACS-CS configurations \cite{PACS-CS:2008bkb}
used in the analysis of the nucleon spectrum in full QCD \cite{Liu:2016uzk}.  We aim to explore the
nucleon spectrum in a quenched simulation with matched lattice spacing and quark masses to evaluate
the sensitivity of the $1s$ and $2s$ state energies observed on the lattice.

Thus, we seek the same $32^3 \times 64$ lattice volume.  We consider the same ${\mathcal
  O}(a^2)$-improved Iwasaki gauge-field action \cite{Iwasaki:1985we} with $\beta$ tuned to provide
a lattice spacing of $a = 0.100$ fm as determined via the Sommer scheme \cite{Sommer:1993ce}.

To access the same light quark masses in the quenched version of the theory, we use the fat-link
clover action with three sweeps of stout-link smearing applied to the links \cite{Moran:2010rn}
with an isotropic smearing parameter $\rho = 0.1$.  With this, the mean link, $u_0 = \langle \tr
\square / 3 \rangle^{1/4} = 0.993$, such that mean-field improvement of the tree-level clover
coefficient, $C_{\rm SW} = 1$ is sufficient to obtain ${\mathcal O}(a)$ improvement of this
Wilson-type action. The use of fat links in the action addresses the exceptional configuration
problem associated with large additive mass renormalisations which otherwise make the simulations
at the lightest quark mass untenable in the quenched theory.
The hopping parameters of this fermion action in the pure gauge theory are tuned to replicate the
pion masses realised in the PACS-CS simulations \cite{PACS-CS:2008bkb}.  These tuned parameters are
summarised in Table \ref{tab:kappas}.

To enhance statistics, we consider multiple fermion sources on each of the 2+1 flavour gauge field
configurations, as summarised in Table \ref{tab:kappas}.  Configurations are circularly shifted in
space and time directions after which a
fixed boundary condition is introduced at $t=N_t=64$.  We place the fermion source away from the
boundary at $t_s=N_t/4=16$ such that masses extracted at large Euclidean time are maximally
displaced from the boundary.  

Our selection of a fixed boundary condition prevents states from wrapping around the lattice.  This
enables a careful examination of the exponential time dependence without significant artefacts.
Only the pion correlator lives long enough to reveal the effect of the fixed boundary condition in
our simulations and this occurs after $t=N_t \,3/4 = 48$ where the pion effective mass
systematically rises just more than one standard deviation above the normal fluctuations observed.
The ground-state nucleon correlator does not survive long enough to see this boundary effect.  No
systematic drift is observed in the correlator prior to signal loss.

To enhance the basis of interpolating fields employed in our variational analysis, gauge-invariant
Gaussian smearing~\cite{Gusken:1989qx} is used at the fermion source and sink with a fixed smearing
fraction and four different smearing levels.  Following the results of the early comprehensive
analyses of Refs.~\cite{Mahbub:2009aa,Mahbub:2009aa}, we consider 16, 35, 100, and 200 smearing
sweeps with a smearing fraction of \(\alpha = 0.7\).  Variational parameters are fixed at $t_0 =
t_s + 1$ and $\Delta t = 2$, except for the second lightest mass, where $\Delta t = 3$ is used
to avoid a degeneracy in the variational analysis.

\section{Excitation properties}

\subsection{Interpolating fields}

\begin{figure}
\centering
\includegraphics[clip=true,trim=0.0cm 0.0cm 0.0cm 0.5cm,width=0.49\linewidth]{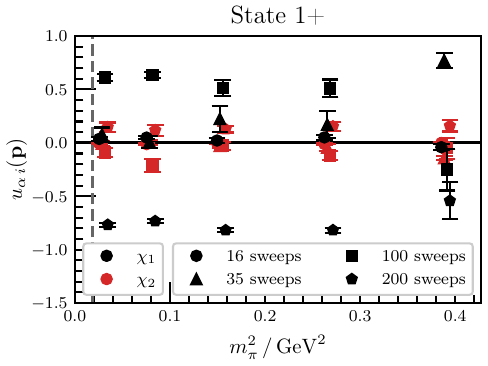}
\includegraphics[clip=true,trim=0.0cm 0.0cm 0.0cm 0.5cm,width=0.49\linewidth]{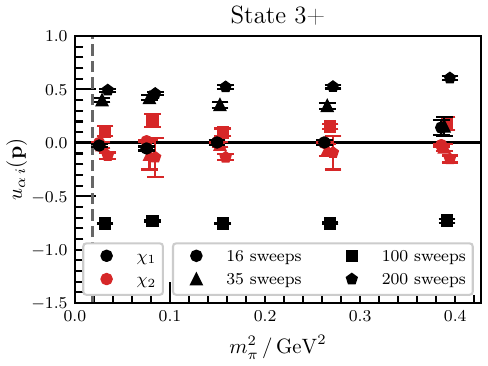}
  \caption{Eigenvector components $u_{\alpha\, i}(\vect p)$ for the $2s$ (left) and $3s$ (right) nucleon
  excitations in full 2+1 flavour QCD at momentum $\vect p = \vect 0$ at the five
  quark masses of the PACS-CS ensembles \cite{PACS-CS:2008bkb}.  Here the index $i$ runs over eight
  values for the two interpolators, $\chi_1$ and $\chi_2$, times the four different gauge invariant
  Gaussian smearing \cite{Gusken:1989qx} levels, 16, 35, 100 and 200 sweeps as discussed in the
  text.  }
\label{fig:evFull}
\end{figure} 

\begin{figure}
\centering
\includegraphics[clip=true,trim=0.0cm 0.0cm 0.0cm 0.5cm,width=0.49\linewidth]{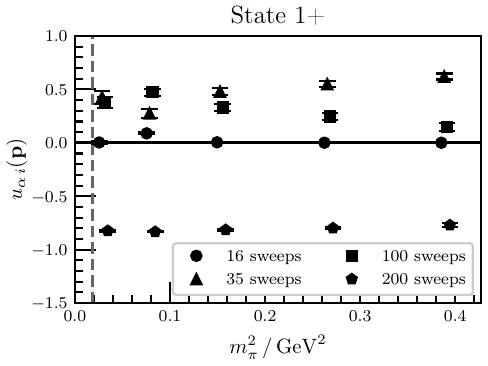}
\includegraphics[clip=true,trim=0.0cm 0.0cm 0.0cm 0.5cm,width=0.49\linewidth]{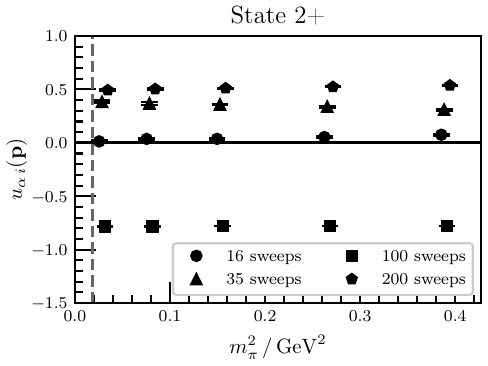}
\caption{Eigenvector components $u_{\alpha\, i}(\vect p)$ for the 2s (left) and 3s (right)
  positive parity nucleon excitations in quenched QCD at momentum
  $\vect p = \vect 0$.  Calculations are performed at $\kappa_Q$ values tuned to reproduce the five
  quark masses of the PACS-CS ensembles \cite{PACS-CS:2008bkb}.  Here the index $i$ runs over the
  four different gauge invariant Gaussian smearing \cite{Gusken:1989qx} levels, 16, 35, 100 and 200
  sweeps for the interpolator, $\chi_1$.  }
\label{fig:evQuen}
\end{figure} 

We begin with an examination of the role of the correlation matrix interpolating fields in exciting
the $2s$ state observed in our full QCD simulations.  Figure~\ref{fig:evFull} illustrates the
eigenvector components $u_{\alpha\, i}(\vect p)$ for the $2s$ (left) and $3s$ (right) nucleon excitations,
at momentum $\vect p = \vect 0$ at the five quark masses of the PACS-CS
ensembles. Here the index $i$ runs over eight values for the two interpolators, $\chi_1$ and
$\chi_2$, times the four different gauge invariant Gaussian smearing levels, 16, 35, 100 and 200
sweeps.

The first feature to note is that the interpolator $\chi_2$ plays a negligible role in exciting
either radial excitation.  It was dropped in the wave function analyses of
Refs.~\cite{Roberts:2013ipa,Roberts:2013oea} with little effect.
The insensitivity of the spectrum obtained with our single-state ansatz is well documented
\cite{Mahbub:2009aa,Mahbub:2009aa,Kiratidis:2015vpa,Kiratidis:2016hda}. While the presence or
absence of a state does indeed depend on the interpolating fields used to form the correlation
matrix, the energy of the observed excitation is not sensitive. 

In the quenched theory, flavour-singlet dressings of hadron states can become problematic at the
lighter quark masses considered.  Upon quenching the theory, the $\eta'$ meson remains degenerate
with the pion, and its two-loop disconnected quark-flow structure generates negative metric
contributions that can render correlation functions that would be positive definite in the full
theory negative.
By dropping $\chi_2$ from the quenched analysis, we are able to minimise coupling to these
problematic contributions.

The second main feature to be drawn from Fig.~\ref{fig:evFull}, is that the $2s$ state's node structure is
predominantly governed through the interplay between the 100 and 200 sweep interpolators.
The familiar superposition of broad and narrow distributions with a relative minus sign
\cite{Mahbub:2013ala,Hockley:2023xms,Hockley:2024aym} forms a single node in the wave function
\cite{Roberts:2013ipa,Roberts:2013oea}, a signature of this state.  It's interesting that at
the heaviest quark mass considered, the central peak is narrower, this time superposing a narrow 35
sweep source with a linear combination of broader 100 and 200 sweep sources with signs opposite the
central 35 sweep source.

Turning now to the $3s$ excitation, we observe sign changes from the narrow 35 sweep source, to
the 100 sweep source and again from the 100 sweep source to the 200 sweep source, thus creating two
nodes in the wave function. Once again
a narrowing is observed in the central peak at the heaviest quark mass, with the 16 sweep source
combining with the 35 sweep source.

Proceeding to the quenched theory, we observe the same general structure where  narrow sources are
superposed with broad sources to create nodes in the wave function.  Figure~\ref{fig:evQuen}
illustrates the eigenvectors for the $2s$ (left) and $3s$ (right) excitations.  

We begin with consideration of the $2s$ excitation.  This time the broad 200 sweep source is
superposed with a liner combination of narrower sources with signs opposite to the broad source.
Again, a single node is produced.  It is interesting to see the more enhanced role for the narrower
35 sweep source at the second-heaviest quark mass.  This state is smaller in the
quenched theory, in accord with a suppression of meson-baryon couplings.  Moreover, the
quark mass dependence of the size of the inner peak, driven by interplay between 35 and 100 sweeps
is in accord with expectations.  At the heaviest quark mass, 35 sweeps plays the
dominate role as in full QCD, but the 100 sweep interpolator comes into play more as the quarks
become light.  Thus, at the lighter quark masses considered, the change in the structure of the
quenched states is more subtle.

The $3s$ excitation displays a very similar set of eigenvectors to the full QCD simulations. The only
notable difference is that there is less narrowing at the heaviest quark mass.
We note that for both states considered here, the 16 sweep source plays
only a small role. In every case its sign matches the 35 sweep source such that there is no
chance of it creating an additional node in the wave function.

\begin{figure}
\centering
\includegraphics[clip=true,trim=0.0cm 0.0cm 0.0cm 0.5cm,width=0.9\linewidth]{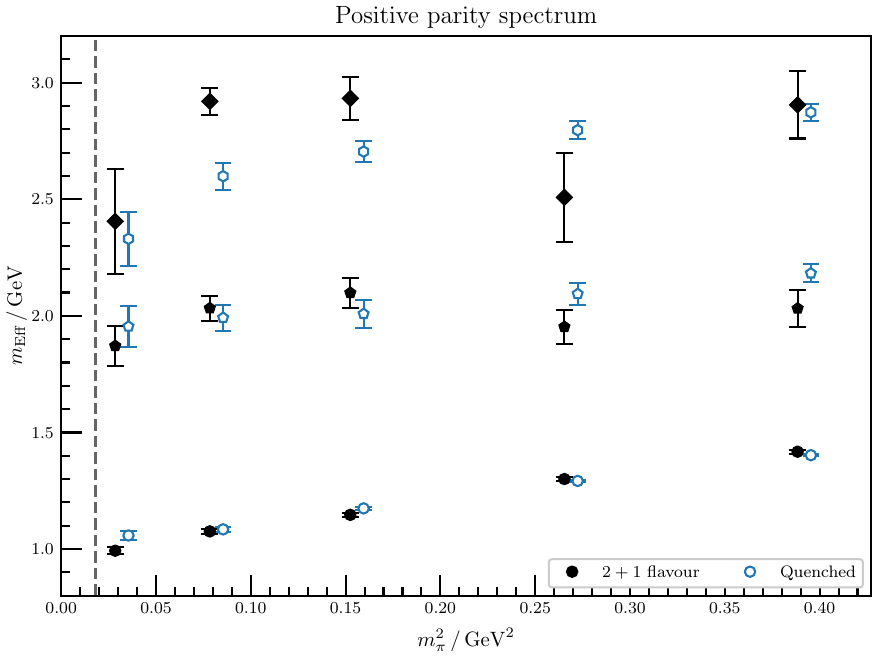}
\caption{
  1s, 2s and 3s nucleon mass spectrum observed in lattice QCD calculations
  in full 2+1 flavour QCD (filled symbols) and quenched QCD (open symbols).  While the
  energy of most excitations change more than $1\sigma$, the three states at $\sim 2$ GeV for the
  lightest three quark masses remain invariant at the $1\sigma$ level upon quenching.  }
\label{fig:spectrum}
\end{figure}

\subsection{Nucleon mass spectrum}

With this understanding of how the correlation matrix is acting to excite the states of interest,
we now turn to a careful examination of the nucleon spectrum. Figure~\ref{fig:spectrum} shows the
ground-state nucleon mass and the first two radial excitations as a function of quark mass for both
full $2+1$ flavour QCD (black points) and for the quenched theory (blue points).

The mass of the ground state nucleon is observed to be insensitive to the change in the meson
dressing of the nucleon at all but the lightest quark mass considered.  Here, the single-particle
dominated state is pushed down in the process of mixing with the more energetic two-particle $\pi
N$ states in full QCD. The suppression of these couplings in the quenched theory suppresses this
mixing, leaving the ground state relatively high.

While the energy of most excited states change more than $1\sigma$, there are three states with
relatively small uncertainties that remain invariant at the $1\sigma$ level upon quenching.  The
three states are at $\sim 2$ GeV at the lightest three quark masses.  These are the $2s$
excitations of the nucleon that we have hypothesised should be insensitive if the $2s$ excitation
is indeed associated with the $N1/2^+(1710)$ and $N1/2^+(1880)$ resonances. The small width of the
$N1/2^+(1710)$ suggests invariance due to the small coupling to meson-baryon channels.  Similarly,
the need for photoproduction to see the $N1/2^+(1880)$ resonance again suggests a diminished
sensitivity to changes in the meson-baryon dressings of the $2s$ state.  This has now been
realised, but only at light quark masses approaching the physical point.

At the two largest quark masses considered, the variation of the spectrum is larger.  Here HEFT can
provide some insight into why the variance is more significant.  HEFT describes the lowest-lying
excitations at the two largest quark masses as having strong mixings with the $\pi N$ two-particle
basis states \cite{Wu:2017qve} as they pass through the lattice QCD points. At the largest quark
mass, the first $\pi N$ two particle basis state is higher in energy than the state observed on the
lattice.  The mixing of the single and two-particle states acts to push down the single particle
dominated state in full QCD.  The process of quenching suppresses the couplings between $N$ and
$\pi N$ and weaker mixing with the higher $\pi N$ basis states causes the quenched results to sit
higher.

\section{Conclusion} 

In summary, we have presented evidence that there are indeed quark-model like states in the nucleon
spectrum. While the positions of the $N1/2^+(1710)$ and $N1/2^+(1880)$ resonances in the nucleon
spectrum are close to the $2s$ excitation observed in lattice QCD, making them plausible
candidates, we now have additional information suggesting these resonances are associated with the
$2s$ excitation.

We have argued the physical properties of these resonances provide additional insight.  The small
width of the $N1/2^+(1710)$ suggests a relatively small coupling to meson-baryon channels such that
the position of the $2s$ state in the spectrum should be largely invariant to changes in the
strength of meson-baryon dressings of the state.  This is echoed by the need for photoproduction to
see the $N1/2^+(1880)$ resonance. Again this property suggests a diminished coupling to
meson-baryon states and thus invariance of the $2s$ state energy upon suppressing the strength of
meson-baryon dressings of the state.

To alter the couplings of two-particle meson-baryon states forming the meson cloud of lattice
energy eigenstates, we have resorted to quenching the theory, removing sea-quark-loop contributions
and thus altering and generally suppressing the meson cloud contributions to eigenstates observed
in quenched lattice calculations.  The anticipated approximate invariance of the $2s$ state seen in
lattice QCD simulations has now been observed at light quark masses approaching the physical point.

The eigenvectors of the lattice variational method describe these radial excitations of quenched
and full QCD in a consistent manner.  The first and second excitations have one and two nodes
respectively.  Moreover, the quark mass dependence of the node radius is in accord with expectations
with heavier quark masses corresponding to smaller radii.  While the quenched states do tend to be
smaller in some cases, the change is subtle at the lightest quark masses as the 100 sweep smearing
interpolator comes into play.

In interpreting and understanding the lattice results, we have drawn on results from Hamiltonian
Effective Field Theory which provides insight into the composition of baryon excitations in terms
of the superposition of basis states required to construct the eigenstates of the Hamiltonian.  Of
particular interest is the prediction of states near 2 GeV dominated by single-particle
contributions \cite{Wu:2017qve} to the extent that their properties are expected to be insensitive
to changes in the meson-baryon couplings.  This expectation has been observed in our lattice QCD
simulations.  At larger quark masses where the first excitation observed in lattice QCD has a
larger $\pi N$ component \cite{Wu:2017qve}, significant movement in the mass of the state is
anticipated and observed.

We note that the masses of these radial excitations in the quenched theory remain far from
the Roper resonance energy further supporting the contemporary perspective that the structure of
the Roper is complicated and beyond the scope of a conventional quark model radial excitation.

Consequentially, accessing finite-volume states in the region of the Roper resonance requires
the consideration of non-local momentum-projected two-particle interpolating fields, similar to
what has been done for the $\Lambda(1405)$~\cite{BaryonScatteringBaSc:2023zvt,BaryonScatteringBaSc:2023ori}. As such,
the determination of the electromagnetic form factors and transition moments of the Roper
demand three-point function techniques on momentum projected five-quark correlation functions,
presenting a significant challenge to the community.

Associating the $2s$ excitation with the $N1/2^+(1710)$ and $N1/2^+(1880)$ resonances also provides
a resolution to the missing baryon resonances problem \cite{Leinweber:2024psf,Hockley:2023yzn}.
Had the parameters of the quark model been tuned to place the $2s$ excitation at $\sim 2$ GeV, the
next excitation would be well above 2 GeV.  In this regard, the missing resonance problem may be
regarded as further evidence that the $2s$ radial excitation of the nucleon is not associated with
the Roper resonance, but rather the $N1/2^+(1710)$ and $N1/2^+(1880)$ resonances.

%

\ack 

We thank the PACS-CS Collaboration for making their configurations available via
the International Lattice Data Grid (ILDG). This research was supported with
supercomputing resources provided by the Phoenix HPC service at
the University of Adelaide. This research was undertaken with the assistance of resources from the
National Computational Infrastructure (NCI), provided through the National Computational Merit
Allocation Scheme, and supported by the Australian Government through Grant No. LE190100021 via the
University of Adelaide Partner Share. This research was supported by the Australian Research
Council through Grants No. DP150103164, DP190102215 and DP210103706 (DBL). WK was supported by the
Pawsey Supercomputing Centre through the Pawsey Centre for Extreme Scale Readiness (PaCER) program.
FS is supported by a Ramsay Fellowship from the University of Adelaide. 
	
\section*{References}
\bibliographystyle{unsrturl}
\bibliography{invariance}
	
\end{document}